\documentclass[showkeywords,onecolumn,amssymb,amsmath,natbib,%
   superscriptaddress]{revtex4}
\usepackage{amssymb}
\usepackage{epsfig}

\begin{document}

\title{ Thermalized Non-Equilibrated Matter: Compound Processes and Beyond}

\author{L. Benet}%
\affiliation{Instituto de Ciencias F\'{\i}sicas,
  Univeridad Nacional Aut\'onoma de M\'exico (UNAM)\\
  Apdo. Postal 48--3, 62251--Cuernavaca, Morelos, M\'exico}%
\author{S. Hern\'andez-Quiroz}%
\affiliation{Facultad de Ciencias, Universidad del Estado de Morelos (UAEM)\\
  62209-Cuernavaca, Morelos, Mexico}%
\author{S.Y. Kun}%
\affiliation{Facultad de Ciencias, Universidad del Estado de Morelos (UAEM)\\
  62209-Cuernavaca, Morelos, Mexico}%
\affiliation{Nonlinear Physics Center, RSPhysSE\\
  The Australian National University (ANU)\\
  Canberra ACT 0200, Australia}%

\date{\today}

\begin{abstract}
  A characteristic feature of thermalized non-equilibrated matter is
  that, in spite of energy relaxation (thermalization), a phase memory
  of the way the strongly interacting many-body system was excited
  remains. In this contribution we analyze a low energy evaporating
  proton data in nucleon induced reactions at $\simeq$62 MeV incident
  energy with $^{197}$Au, $^{208}$Pb, $^{209}$Bi and $^{nat}$U. Our
  analysis demonstrates that the thermalized non-equilibrated matter
  survives a cascade of several evaporating particles. Thus the
  experiments show that the effect of the anomalously slow phase
  relaxation, with upper limits of the phase relaxation widths in the range
  1-10$^{-4}$ eV, is stable with respect to the multi-step evaporating cascade 
  from the thermalized compound nuclei. We also briefly mention
  manifestations and implications of the thermalized non-equilibrated
  matter for some other fields.
\end{abstract}

\maketitle

\section{Introduction}
\label{sec1}

The basic idea of the modern theory of highly excited strongly
interacting quantum many-body systems is that the phase relaxation
time is much shorter than the energy relaxation
time~\cite{GMGW1998}. Application of this idea to the theory of
pre-compound and compound reactions implies the absence of
correlations between energy fluctuating around zero S-matrix elements
carrying different total spin and parity quantum numbers. However the
assumption of a very fast phase relaxation is in a conflict with many
data sets on complex quantum collisions. In particular, the
anomalously long-lived spin off-diagonal S-matrix correlations have
been revealed from the data on forward peaking of evaporating
protons in nucleon induced~\cite{FKS2005,BFK2006} and
photonuclear~\cite{BFKS2006,BBK2007} reactions. These long-lived
correlations reflect an anomalously slow phase relaxation, which is
many orders of magnitude longer than the energy relaxation. This
provides a manifestation of a new form of matter: thermalized
non-equilibrated matter introduced by one of
us~\cite{K1993,K1994,K1997a}. The problem is of a primary importance
for the nuclear technology applications.

The previous analysis~\cite{FKS2005,BFK2006,BFKS2006,BBK2007} dealt
with the data obtained for relatively low, $\simeq$15-20 MeV,
excitation energies of the compound nuclei. For such energies a second
chance proton evaporation is either forbidden or negligible. In this
contribution we analyze the data sets for the higher energy when there
is a high probability for evaporation up to three and even four
nucleons from the thermalized compound nucleus. The data demonstrate
that, after evaporation of nucleons, the residual compound nuclei are
in (i)~a coherent superposition of the strongly overlapping resonance
states with different total spins and parities, and (ii)~the phase
relaxation between the strongly overlapping resonance states with
different total spins and parities is still anomalously long in these
highly-excited residual compound nuclei.

\section{Analysis of the data}
\label{sec2}

In Figs.~1-6 we present the scaled low energy proton spectra, for
forward and backwards angles, produced in nucleon induced reactions
with the incident energy $\simeq$62 MeV. The spectra are in arbitrary
units and are scaled with (i)~the outgoing proton energy $\varepsilon$, 
and (ii)~the
cross section, $P(\varepsilon )$, of the inverse process of the capture 
of the proton with
 energy $\varepsilon$ by the residual compound nucleus. The
latter quantities are taken from~\cite{S1953} for $r_0$=1.3 fm.

From the comprehensive analysis~\cite{KA1993} of the $^{209}$Bi(p,p')
scattering at E$_{p}$=62 MeV it follows that the multi-step direct
reaction contribution is negligible for E$_{p'}\leq 10$ MeV.  From the
structure of the multi-step direct reactions models it is clear that
this should apply to all the data sets analyzed in this contribution.
Our analysis, based on the exiton model~\cite{GMGW1998}, shows that,
for E$_{p'}\leq 10$ MeV, the multi-step compound reaction contribution
is also negligible for all the data sets analyzed here (Figs.~1-6).
Thus, for E$_{p'}\leq 10$ MeV, we deal exclusively with evaporation
cascade from the chain of thermalized cooling down compound
nuclei. Indeed the scaled spectra show locally exponential $\varepsilon$-dependence,
in agreement with the statistical evaporation
model~\cite{GMGW1998}. The slopes of the scaled spectra --- the inverse
nuclear ``temperatures'' --- decrease with decrease of the energy of
the evaporated protons. This reflects a process of cooling down of the
compound nuclei by means of particle evaporation: the smallest
energies of the evaporating protons correspond to the mainly last
chance of evaporation. Notice that the slightly different slopes
for forward and backward angles do not mean  different ``temperatures''
but reflect different $\varepsilon$-dependence of the transmission coefficients
for different exit channel orbital momenta of evaporating protons.
\begin{figure}[!b]
\includegraphics[width=100mm,height=100mm,angle=0]{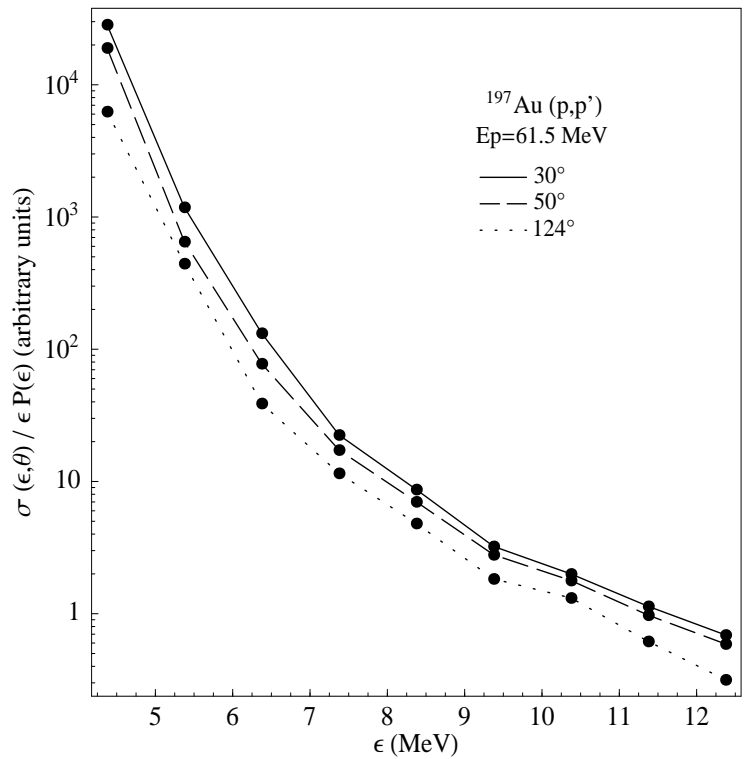}
\caption{\label{f1}%
The scaled outgoing proton spectra, for different angles, from
the $^{197}$Au(p,p') reaction. The lines connecting the data full
points are for eye guide. The data errors reported are about of the
size of the full dots.  The data are from~\cite{BP1973}.
 }
\end{figure}
\begin{figure}
\includegraphics[width=100mm,height=100mm,angle=0]{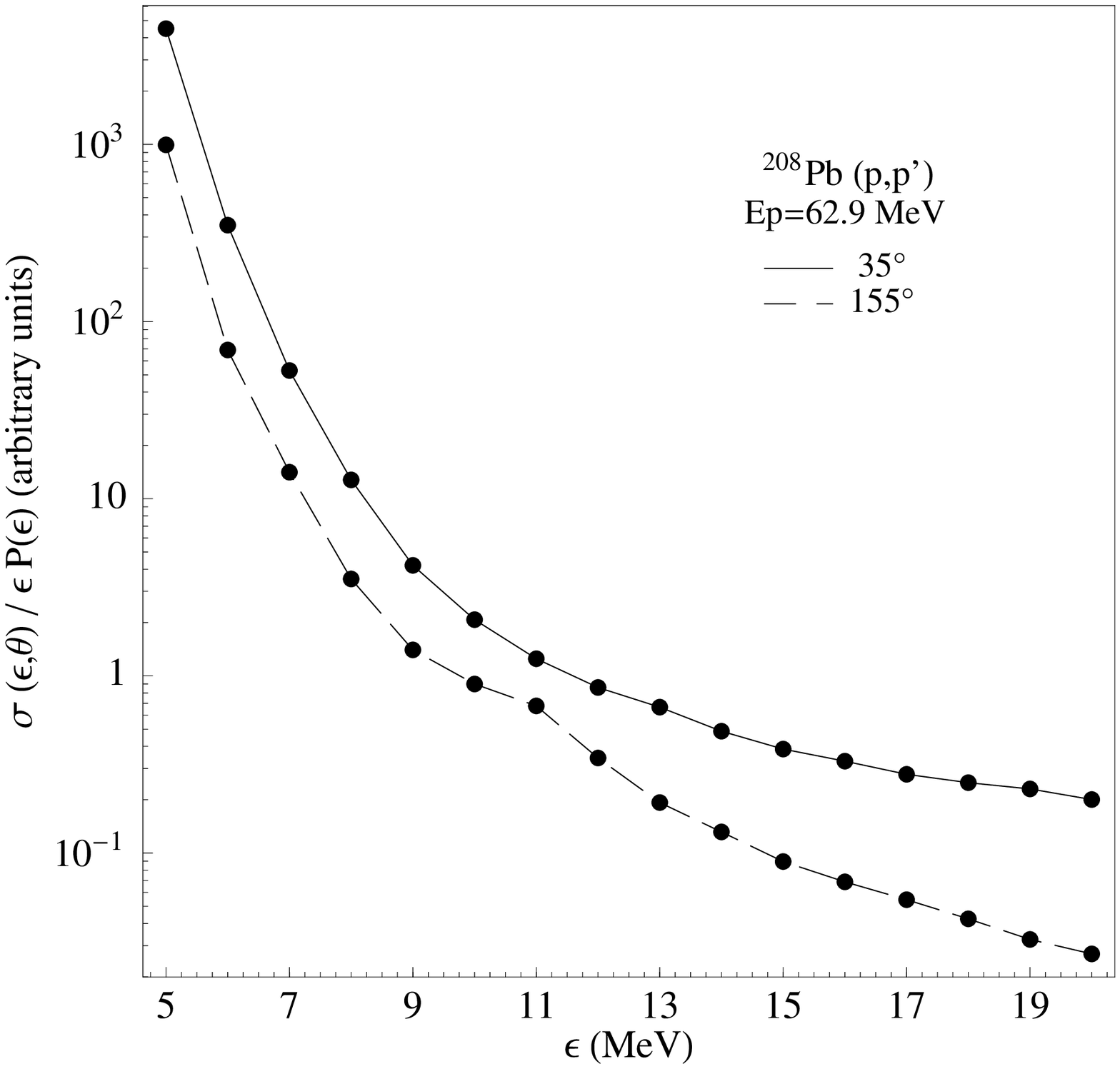}
\caption{\label{f2}%
The same as in Fig. 1 but for the $^{208}$Pb(p,p') reaction.
The data are from~\cite{G2005}.
 }
 \end{figure}
\begin{figure}
\includegraphics[width=100mm,height=100mm,angle=0]{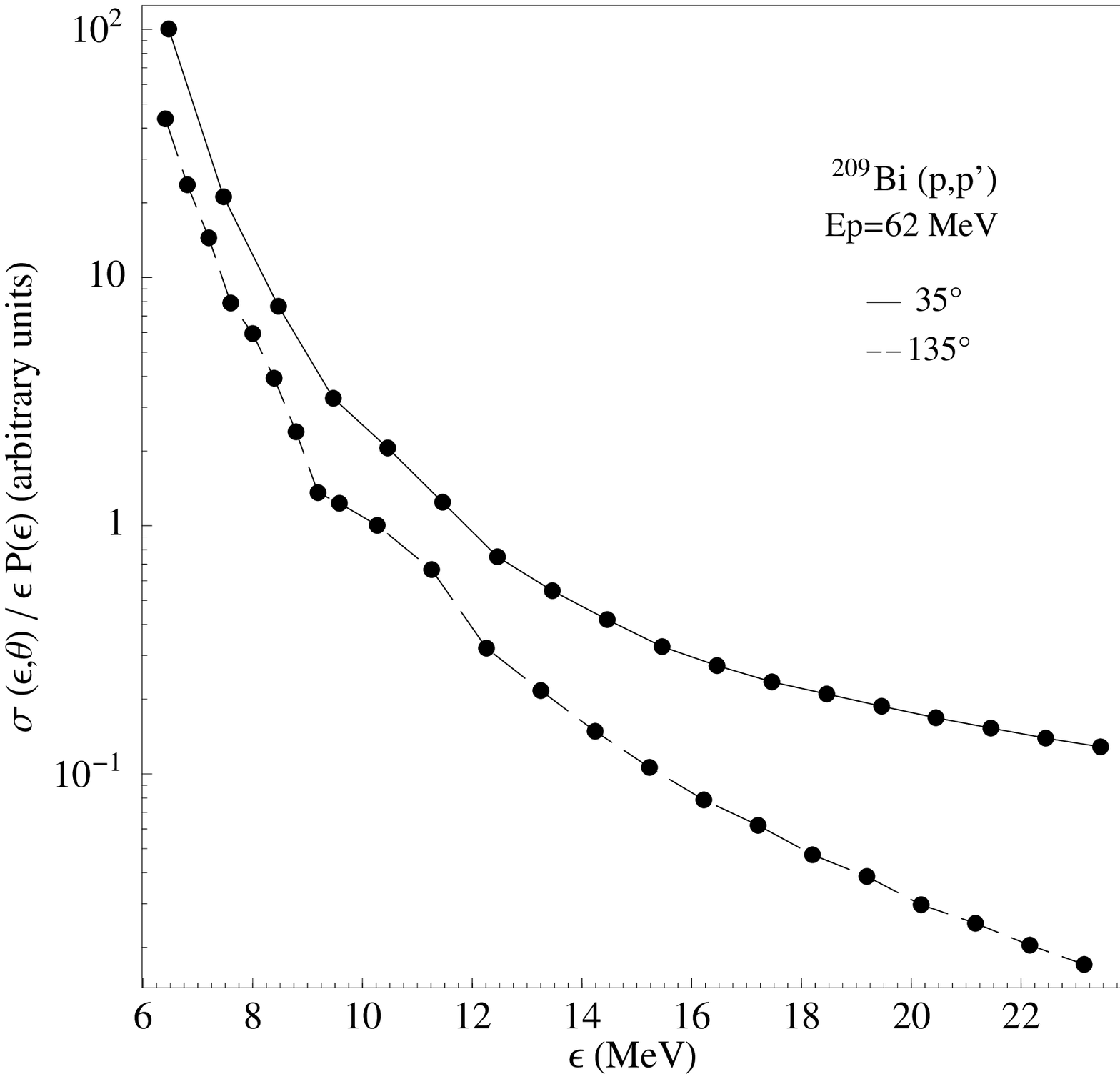}
\caption{\label{f3}%
The same as in Fig.~1 but for the $^{209}$Bi(p,p')
reaction. The data are from~\cite{BP1973}.
 }
\end{figure}
\begin{figure}
\includegraphics[width=100mm,height=100mm,angle=0]{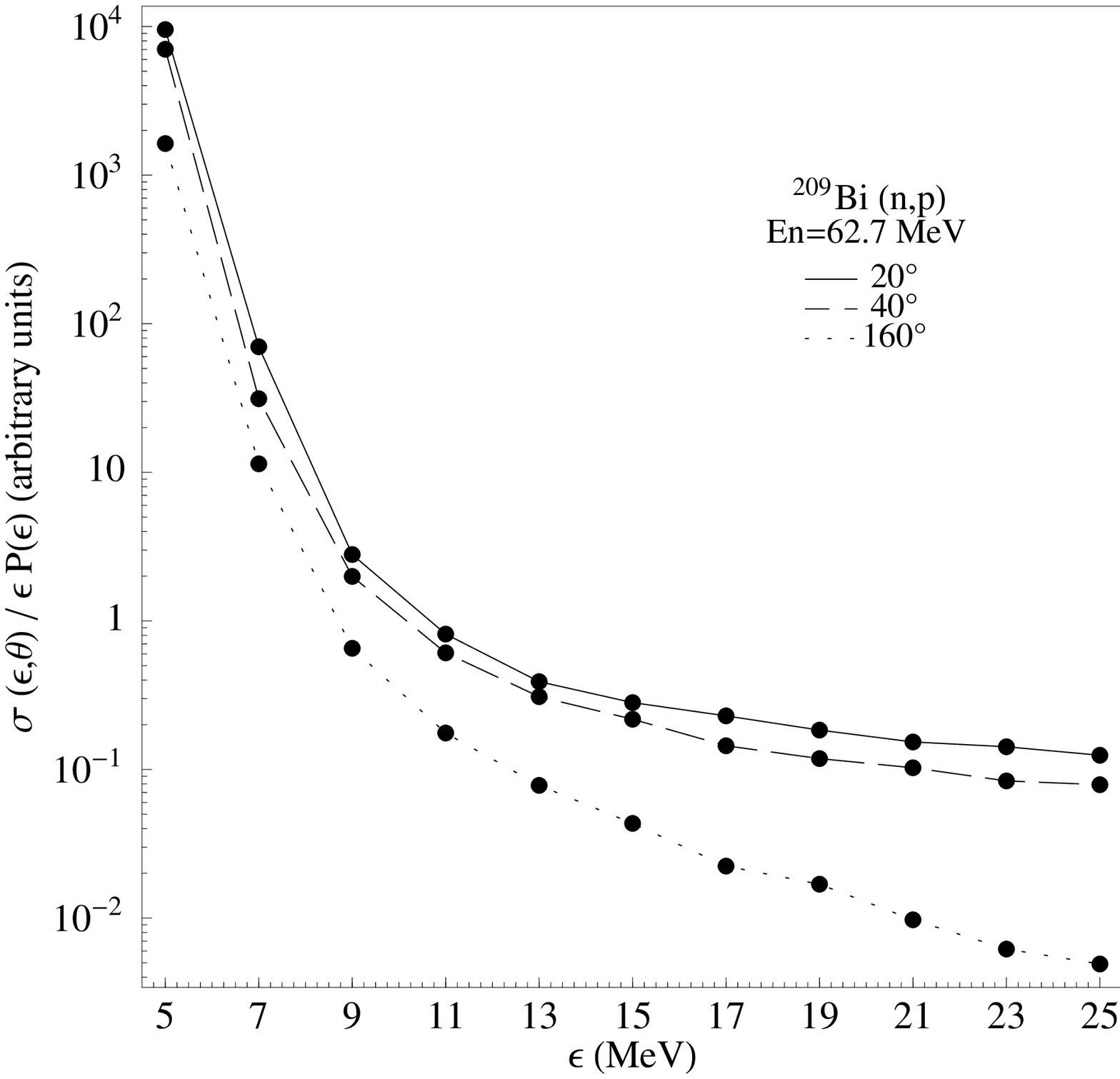}
\caption{\label{f4}%
The same as in Fig. 1 but for the $^{209}$Bi(n,p) reaction.
The data are from~\cite{Rea2003a}.
 }
 \end{figure}
\begin{figure}
\includegraphics[width=100mm,height=100mm,angle=0]{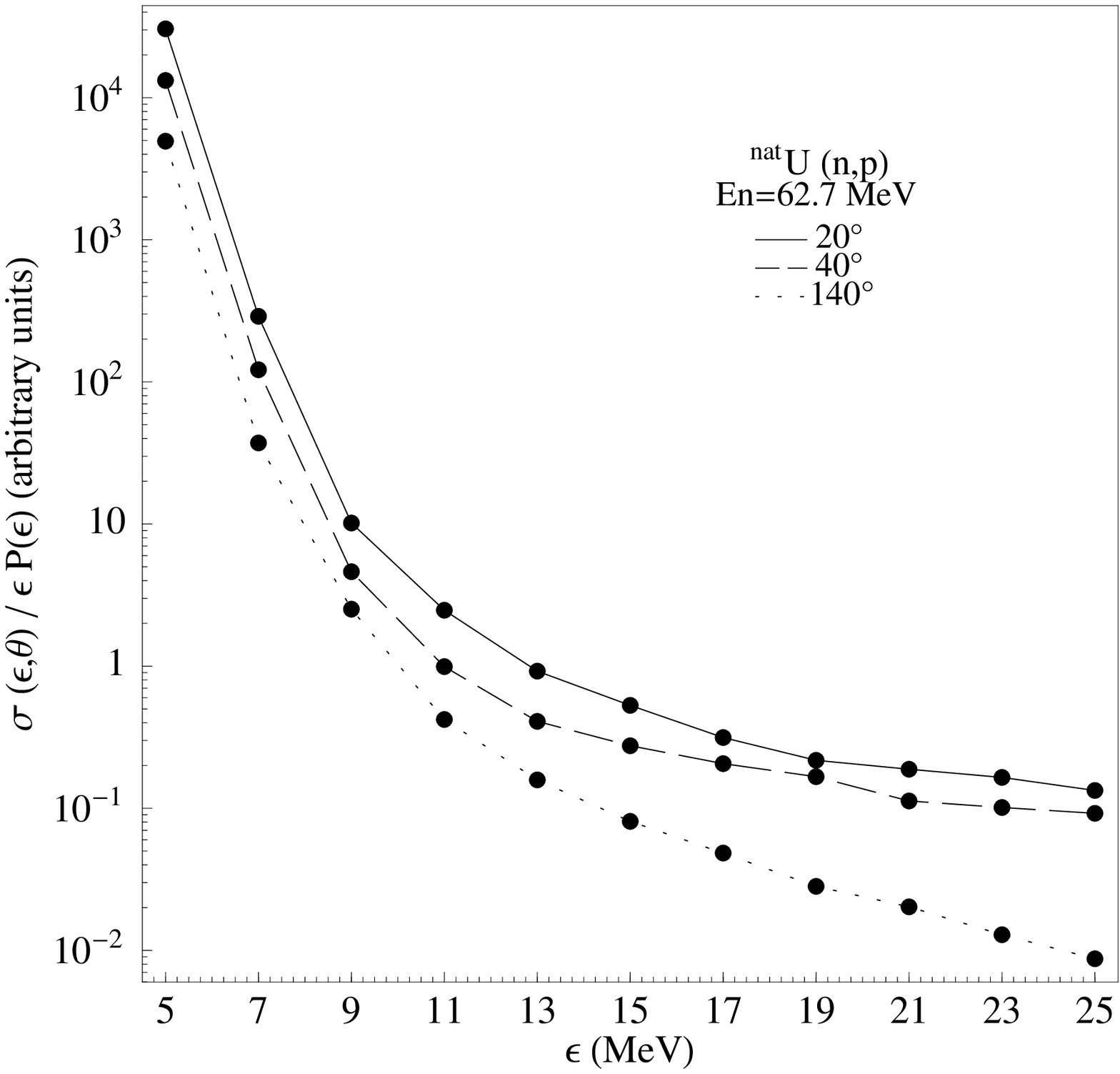}
\caption{\label{f5}%
The same as in Fig. 1 but for the $^{nat}$U(n,p) reaction.
The data are from~\cite{Rea2003b}.
}
\end{figure}
\begin{figure}
\includegraphics[width=100mm,height=100mm,angle=0]{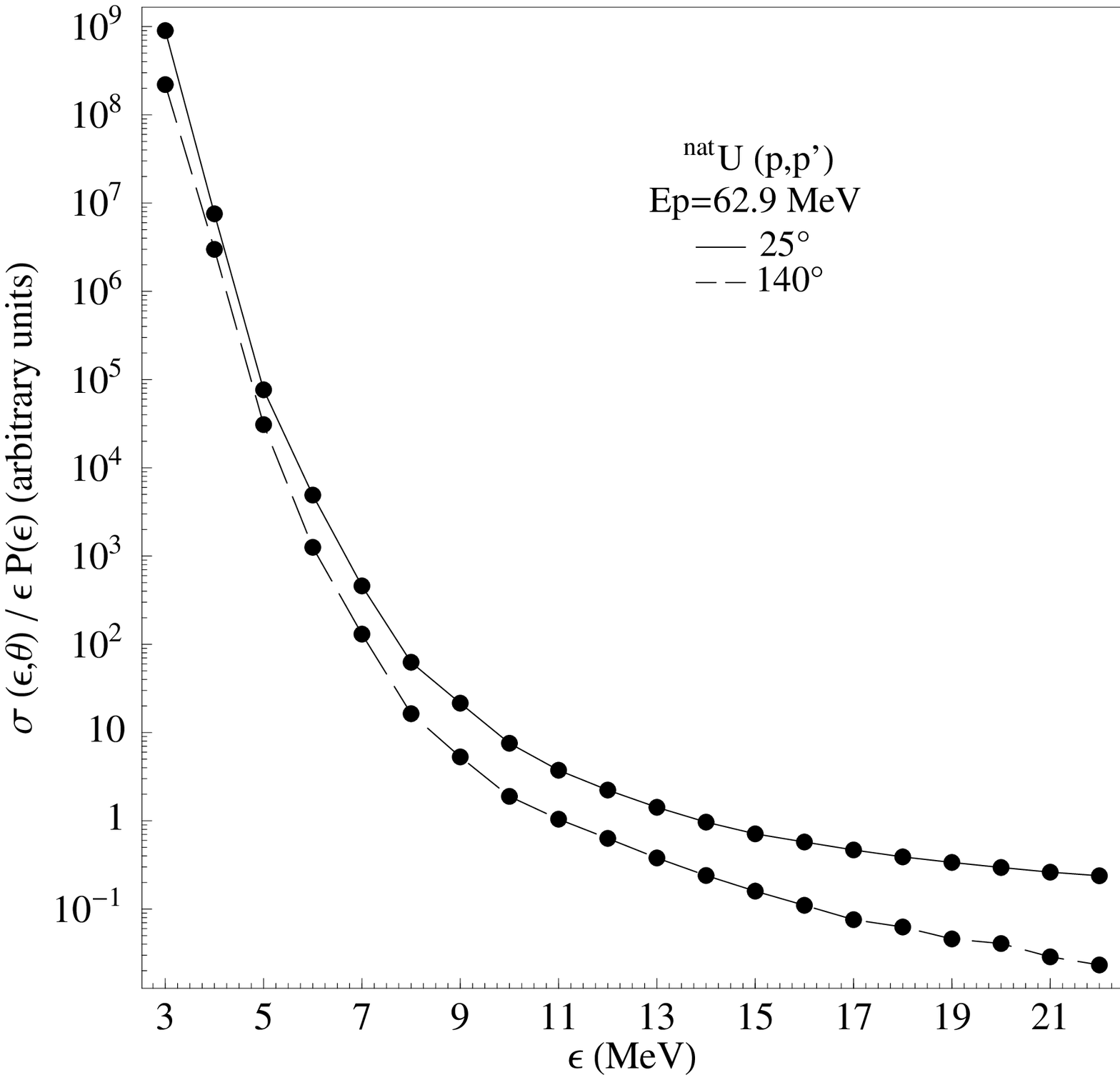}
\caption{\label{f6}%
The same as in Fig.~1 but for the $^{nat}$U(p,p') reaction.
The data are from~\cite{G05arXiv}. 
}
\end{figure}
Yet, in spite of a complete energy relaxation (thermalization) one observes that
angular distributions of the evaporating protons are not symmetric
around 90$^\circ$ up to the last chance thermal proton emission. The
forward thermal emission intensities exceed the backward ones by factors of
2 to 6. The data~\cite{BP1973,G2005,Rea2003a,Rea2003b,G05arXiv} are in
good agreement with each other and show that forward peaking is
basically present for many angular distributions of $d$, $t$, $^3$He
and $\alpha$-particles in the low energy, typically evaporation,
domain of the outgoing spectra. In Ref.~\cite{G2005}, a forward
peaking of evaporating neutrons is also reported. Thus, the data in
Figs.~1-6 demonstrate that thermalized non-equilibrated matter is
stable with respect to a multi-step evaporation cascade. This also means that
the phase relaxation widths $\beta$, introduced by one of
us~\cite{K1993,K1994,K1997a}, are about or less than compound nucleus
total decay widths (defined as inverse average compound nuclei
life-times), on each step of the cascade. The smallest
compound-nucleus total-decay width corresponds to the last chance
evaporation, i.e. to the smallest proton outgoing energy. From the slopes
of the scaled spectra at the smallest energies in Figs.~1-6 we
evaluate the ``temperatures'' of the residual compound nuclei. This
allows us to evaluate excitation energies of the compound nuclei from
which protons, with the lowest measured energies in Figs.~1-6, were
evaporated. Finally, using the statistical model expression for the
compound nucleus total decay width, which is in a good agreement with
the data systematics~\cite{EMK1966}, we find the upper limits for the
phase relaxation widths for the data in Figs.~1-6. These upper limits
are in the range from the maximal value of $\simeq$1 eV for the
data~\cite{BP1973} (Fig.~3) (because of the lowest energy
reported~\cite{BP1973} is maximal among all the other data sets
analyzed here) down to the value of $\simeq$10$^{-4}$ eV for the
data~\cite{G05arXiv}(Fig. 6). For a comparison, typical nuclear energy
relaxation widths are evaluated as $\Gamma_{spr}\geq$1 MeV~\cite{GMGW1998}. 
We note that
in all the data sets in Figs.~1-6, the excitation energy of the
compound nuclei for the last chance proton evaporation is $\geq$ 10
MeV. This means that the average level-spacing $D\leq 10^{-16}$ MeV
yielding effective dimensions of Hilbert spaces $\Gamma_{spr}/D\geq
10^{16}$.  Thus, the effect of anomalously slow phase relaxation for
the data sets analyzed in this contribution is again revealed for
exponentially large effective dimensions of Hilbert
space~\cite{K1997a,FKS2005,BFKS2006}.  Yet, for the upper limit of
lowest $\beta\simeq 10^{-4}$ eV obtained above, we have $\beta
/D\simeq 10^{6}$.

\section{Discussion and conclusion}
\label{sec3}

We briefly mention that the new form of matter --- thermalized
non-equilibrated matter --- is of primary importance for many-qubit
quantum computation since anomalously long-lived phase memory can
extend the time for quantum computation far beyond the quantum chaos
border~\cite{FKS2005,BFKS2006}. An effect of a very slow phase
relaxation has also been strongly supported by numerical calculations
for H+D$_2$~\cite{Aea2002}, F+HD~\cite{A2003} and
He+H$^+_2$~\cite{PA2006} chemical reactions. In these calculations, a
slow phase relaxation ($\beta\ll\Gamma_{spr}$) manifests itself in
stable rotating wave-packets of the excited intermediate
complexes~\cite{BCKW2007}. Interestingly, this same effect of stable
coherent rotation~\cite{K1994b,K1997b} was originally revealed and
described for heavy-ion elastic and inelastic scattering, e.g. for
$^{12}$C+$^{24}$Mg~\cite{KVV1999,KVG2001}, $^{24}$Mg+$^{24}$Mg and
$^{28}$Si+$^{28}$Si~\cite{KRV1999}, $^{58}$Ni+$^{46}$Ti and
$^{58}$Ni+$^{62}$Ni~\cite{Kea1997a} scattering (see
also~\cite{Bea2008}) as well as for strongly dissipative heavy-ion
collisions, e.g., $^{19}$F+$^{89}$Y~\cite{Kea1997b},
$^{28}$Si+$^{64}$Ni~\cite{KDV1997} and
$^{28}$Si+$^{48}$Ti~\cite{KN1992,KNP1993} collisions.  Slow phase
relaxation has been found responsible for quantum-classical transition and 
Schr\"odinger cat states
 in heavy-ion collisions and chemical
reactions~\cite{Kea2003,BKWD2005,BKW2006,BCKW2007}.

Another manifestation of the thermalized non-equilibrated matter is a
strong channel-channel correlation in dissipative heavy-ion
collisions~\cite{K2001}. The correlation has nothing to do with S-matrix unitary 
condition for fixed total spin and parity quantum numbers but originates
from the correlation and slow phase relaxation between the strongly overlapping 
resonance states carrying different total spin and parity values.
This correlation results in
non-self-averaging of excitation function oscillations in these
processes. This signifies that $\beta$ values are much smaller than
$\Gamma_{spr}$ and yet requires finite $\beta$ values, comparable with
the total decay width of the intermediate system~\cite{K2001}. Recently, 
 a fine energy resolution measurement of excitation functions for
$\alpha$-particle yields in $^{19}$F(110-118 MeV)+$^{27}$Al with the
energy step 250 keV in laboratory system has been performed~\cite{WKup}. For each
incident energy, the $\alpha$-particle energy spectra have typical
evaporation shape. Yet, in spite of a summation over the whole evaporation spectra
for each energy step, the corresponding excitation
functions show strong oscillations. This provides yet another demonstration that even though
the intermediate system is thermalized it is not
equilibrated~\cite{K1999}. Thermalization does not guarantee
equilibrium!

There exists an indication that the anomalously slow phase relaxation and the
thermalized non-equilibrated matter may lead to anomalous sensitivity of
the cross sections in complex collisions~\cite{K2000}. The phenomenon
is intimately related to the channel-channel correlation and
non-self-averaging of the excitation function oscillations indicating
deterministic randomness in complex collisions. So far the 
experiments~\cite{Wea2003,Wea07arXiv} have been supporting the theoretical
conjecture~\cite{K2000}.

The data analyzed in this contribution demonstrate that, after
the evaporation cascade, the residual compound nuclei are in (i)~a
coherent superposition of the strongly overlapping resonance states
with different total spins and parities, and (ii)~the phase relaxation
between the strongly overlapping resonance states with different total
spins and parities in the highly-excited residual compound nuclei is
still anomalously long corresponding to the upper limit of the phase
relaxation widths to be in the range 1-10$^{-4}$ eV. This is from 6 to
10 orders of magnitude less than typical nuclear energy relaxation (spreading)
widths~\cite{GMGW1998}.

A possibly interesting aspect of the present discussion is the
following. After emitting a first proton the highly excited residual
nucleus is entangled with this proton. The thermally emitted proton is expected
to readily interact with, say, atoms in the target, i.e. with
``environment''. Therefore, this emitted proton very quickly (possibly 
quicker than the average time to complete the whole evaporation
cascade) experiences decoherence. Accordingly, the reduced density
matrix of the residual nucleus may also be expected to
correspond to a fully mixed state unless it belongs to a decoherence
free subspace of the highly excited strongly interacting
many-body system. On the other hand, for the all data sets analyzed in
this paper, it is not excluded that there is a considerable
contribution of events when the proton is emitted on the last stage of
the cascade only, while all the previous cascade steps proceeded
through evaporation of neutrons. Then, since effective interaction of
neutrons with the ``environment'' is much weaker than that of protons,
the decoherence effects should be expected to be suppressed/postponed.
It could be of interest to design experiments to try to find out which
of the two possibilities may actually be realized, leading to the residual
nucleus being in coherent superposition of strongly overlapping
resonance states with different total spin and parity values (not to
mention anomalously long phase relaxation of such a coherent
superposition). One does not a priory exclude a possible relationship between
the robust, against decoherence, quantum superpositions and their 
anomalously slow phase relaxation
in a view that the latter was introduced by means of ensemble averaging
over couplings of a given (fixed) intermediate system with continuum (``environment''). 

\begin{acknowledgments}
  We are grateful to Erwin Raeymackers for kindly providing us with
  the data tables from Refs~\cite{Rea2003a,Rea2003b}. We are thankful
  to Thomas Seligman for useful discussions. We acknowledge financial
  support from the projects IN-111607 (DGAPA-UNAM) and 79988
  (CONACyT). Saul Hern\'andez-Quiroz acknowledges a CONACyT PhD
  scholarship.
\end{acknowledgments}


\begin{thebibliography}{99}

\bibitem{GMGW1998}%
  T. Guhr T., A. M\"uller-Groeling, H.A. Weidenm{\" u}ller,
  Phys. Rep. {\bf 299}, 189 (1998), and references therein.

\bibitem{FKS2005}%
  J. Flores, S.Yu. Kun, and T.H. Seligman, Phys. Rev. E {\bf 72},
  017201 (2005); quant-ph/0502050.

\bibitem{BFK2006}%
M. Bienert, J. Flores, and S.Yu. Kun, Phys. Rev. C {\bf
    74}, 027602 (2006); nucl-ex/0508020.

\bibitem{BFKS2006}%
  M. Bienert, J. Flores, S.Yu. Kun, and T.H. Seligman, Symmetry,
  Integrability and Geometry: Methods and Applications (SIGMA) {\bf
    2}, Paper 027 (2006); quant-ph/0602224.

\bibitem{BBK2007}%
  L. Benet, M. Bienert, S.Yu. Kun, Rad. Effects and Defects in Solids
  {\bf 162} 605 (2007).

\bibitem{K1993}%
  S.Yu. Kun, Phys. Lett. B {\bf 319}, 16 (1993).

\bibitem{K1994}%
  S.Yu. Kun, Z. Phys. A {\bf 348}, 273 (1994).

\bibitem{K1997a}%
  S.Yu. Kun, Z. Phys. A {\bf 357}, 255 (1997).

\bibitem{S1953}%
  M.M. Shapiro, Phys. Rev. {\bf 90}, 171 (1953).

\bibitem{KA1993}%
  A.J. Koning, J.M. Akkermans,  Phys. Rev. C {\bf 47}, 724 (1993).

\bibitem{BP1973}%
  F.E. Bertrand, R.W. Peelle Phys. Rev. C {\bf 8}, 1045 (1973); ORNL
  Report No. ORNL-4460, 1969 (unpublished); ORNL Report No. ORNL-4638,
  1971 (unpublished).

\bibitem{G2005}%
  A. Guertin et al., Eur. Phys. J. A {\bf 23} 49 (2005).

\bibitem{Rea2003a}%
  E. Raeymackers et al., Nucl. Phys. A {\bf 726} 210 (2003).

\bibitem{Rea2003b}%
  E. Raeymackers et al., Phys. Rev. C {\bf 68} 024604 (2003). 

\bibitem{G05arXiv}%
  A. Guertin et al., arXiv:nucl-ex/0512022.

\bibitem{EMK1966}%
  T. Ericson and T. Mayer-Kuckuk, Ann. Rev. Nucl. Sci. {\b 16}, 183
  (1966).

\bibitem{Aea2002}%
  S.C. Althorpe, F. Fern\'andez-Alonso, B.D. Bean, J.D. Ayers,
  A.E. Pomerantz, R.N. Zare, and E. Wrede, Nature (London) {\bf 416},
  67 (2002).

\bibitem{A2003}%
  S.C. Althorpe, J. Phys. Chem. A {\bf 107}, 7152 (2003).

\bibitem{PA2006}%
  A. N. Panda and S.C. Althorpe, Chem. Phys. Lett. {\bf 419}, 245
  (2006).

\bibitem{BCKW2007}%
  L. Benet, L.T. Chadderton, S.Yu. Kun, and Wang Qi, Phys. Rev. A {\bf
    75} 062110 (2007); quant-ph/0610091.

\bibitem{K1994b}%
  S.Yu. Kun, Europhys. Lett. {\bf 26}, 505 (1994).

\bibitem{K1997b}%
  S.Yu. Kun, Z. Phys. A {\bf 357}, 271 (1997).

\bibitem{KVV1999}%
  S.Yu. Kun, A.V. Vagov, and O.K. Vorov, Phys. Rev. C {\bf 59}, R585
  (1999).

\bibitem{KVG2001}%
  S.Yu. Kun, A.V. Vagov, and W. Greiner, Phys. Rev. C {\bf 63}, 014608
  (2001).

\bibitem{KRV1999}%
  S.Yu. Kun, B.A. Robson, and A.V. Vagov, Phys. Rev. Lett. {\bf 83},
  504 (1999).

\bibitem{Kea1997a}%
  S.Yu. Kun {\sl et al.}, Z. Phys. A{\bf 359}, 145 (1997).

\bibitem{Bea2008}%
  L. Benet, L.T. Chadderton, S.Yu. Kun, O.K. Vorov, Q. Wang,
  Phys. At. Nucl. {\bf 71} 819 (2008).

\bibitem{Kea1997b}%
  S.Yu. Kun {\sl et al.}, Z. Phys. A{\bf 359}, 263 (1997).

\bibitem{KDV1997}%
  S.Yu. Kun, V.Yu. Denisov, and A.V. Vagov, Z. Phys. A{\bf 359}, 257
  (1997).

\bibitem{KN1992}%
  S.Yu. Kun, W. N\"orenberg, Z. Phys. A {\bf 343} 215 (1992).

\bibitem{KNP1993}%
  S.Yu. Kun, W. N\"orenberg, M. Papa Phys. Lett. B {\bf 298} 273
  (1993).

\bibitem{Kea2003}%
  S.Yu. Kun, L. Benet, L.T. Chadderton, W. Greiner, and F. Haas,
  Phys. Rev. C {\bf 67}, 011604(R) (2003); quant-ph/0205036.

\bibitem{BKWD2005}%
  L. Benet, S.Yu. Kun, Wang Qi, and V. Denisov, Phys. Lett. B {\bf
    605}, 101 (2005); nucl-th/0407029.

\bibitem{BKW2006}%
  L. Benet, S.Yu. Kun, and Wang Qi, Phys. Rev. C {\bf 73}, 064602
  (2006); quant-ph/0503046.

\bibitem{K2001}%
  S.Yu. Kun, in Non-Equilibrium and Nonlinear dynamics in Nuclear and
  Other Finite Systems, edited by Zhuxia Li, Ke Wu, Xizhen Wu, Enguang
  Zhao, Fumihiko Sakata, AIP Conf. Proc. No. 597 (AIP, Melville, NY,
  2001), p. 319, (2001).

\bibitem{WKup}%
  Wang Qi, S.Yu. Kun, unpublished.

\bibitem{K1999}%
  S.Yu. Kun Phys. Lett. B {\bf 448}, 163 (1999).

\bibitem{K2000}%
  S.Yu. Kun Phys. Rev. Lett. {\bf 84} 423 (2000), and references
  therein.

\bibitem{Wea2003}%
  Wang Qi et al., Int. J. Mod. Phys. E {\bf 12} 377 (2003).

\bibitem{Wea07arXiv}%
  Wang Qi et al., arXiv:quant-ph/0705.4502.

\end{thebibliography}
\end{document}